\documentclass[conference,lettersize]{IEEEtran}
\usepackage{amsmath,amsfonts,amssymb}
\usepackage{physics,amsmath}
\usepackage{algorithm}
\usepackage{algpseudocode}
\let\labelindent\relax
\usepackage{enumitem}

\usepackage{array}
\usepackage{bm}
\usepackage{mathtools}
\usepackage{xcolor}

\usepackage{textcomp}
\usepackage{stfloats}
\usepackage{lipsum}  
\usepackage{url}
\usepackage{verbatim}
\usepackage{graphicx}
\usepackage{subfig}
\usepackage{cite}
\usepackage{pdfpages}
\newcommand*\squeezespaces[1]{
  \thickmuskip=\scalemuskip{\thickmuskip}{#1}%
  \medmuskip=\scalemuskip{\medmuskip}{#1}%
  \thinmuskip=\scalemuskip{\thinmuskip}{#1}%
  \nulldelimiterspace=#1\nulldelimiterspace
  \scriptspace=#1\scriptspace
}
\newcommand*\scalemuskip[2]{%
  \muexpr #1*\numexpr\dimexpr#2pt\relax\relax/65536\relax
}
\setlist[itemize]{leftmargin=*}

\renewcommand\algorithmicthen{}

\DeclareMathOperator{\EX}{E}

\newcommand{\matr}[1]{\mathbf{#1}}
\newcommand{\vect}[1]{\mathbf{#1}}
\DeclareMathOperator{\erfc}{erfc}
\algnewcommand{\IfThenElse}[3]{
  \State \algorithmicif\ #1\ \algorithmicthen\ #2\ \algorithmicelse\ #3}

\begin{document}
\bstctlcite{IEEEexample:BSTcontrol}

\title{Frequency-Domain Detection for Molecular Communications \\
	{\normalfont\large 
		\textbf{Meltem~Civas}\IEEEauthorrefmark{1}\IEEEauthorrefmark{2}\hspace{2mm}\textbf{Ali~Abdali}\IEEEauthorrefmark{1}\hspace{2mm}\textbf{Murat~Kuscu}\IEEEauthorrefmark{1}\hspace{2mm}\textbf{Ozgur~B.~Akan}\IEEEauthorrefmark{1}\IEEEauthorrefmark{2}
	}\\[-1ex]
}
\author{ 
	\IEEEauthorblockA{
		\hspace{0cm}\IEEEauthorrefmark{1}Center for neXt-generation Communications~(CXC)\\ \hspace{0cm}Department of Electrical and Electronics Engineering\\
		\hspace{0cm} Ko\c{c} University, 34450, Istanbul, Turkey\\
		\hspace{0cm} \{mcivas16, aabdali21,  mkuscu, akan\}@ku.edu.tr\hspace{0cm} 
	}\and
	\IEEEauthorblockA{%
		\hspace{0cm}  \IEEEauthorrefmark{2}Internet of Everything (IoE) Group\\ \hspace{0cm}
		Electrical Engineering Division, Department of Engineering\\
		\hspace{0cm} University of Cambridge, CB3 0FA Cambridge, UK \hspace{0cm}\\
		\hspace{0cm} \{mc2365, oba21\}@cam.ac.uk \\
	}
}

\maketitle
\thispagestyle{plain}
\pagenumbering{gobble}
\pagestyle{plain}
\begin{abstract}
Molecular Communications (MC) is a bio-inspired communication paradigm which uses molecules as information carriers, thereby requiring unconventional transmitter/receiver architectures and modulation/detection techniques. Practical MC receivers~(MC-Rxs) can be implemented based on field-effect transistor biosensor (bioFET) architectures, where surface receptors reversibly react with ligands, whose concentration encodes the information. The time-varying concentration of ligand-bound receptors is then translated into electrical signals via field-effect, which is used to decode the transmitted information. However, ligand-receptor interactions do not provide an ideal molecular selectivity, as similar types of ligands, i.e., interferers, co-existing in the MC channel can interact with the same type of receptors, resulting in cross-talk. Overcoming this molecular cross-talk with time-domain samples of the Rx’s electrical output is not always attainable, especially when Rx has no knowledge of the interferer statistics or it operates near saturation. In this study, we propose a frequency-domain detection~(FDD) technique for bioFET-based MC-Rxs, which exploits the difference in binding reaction rates of different types of ligands, reflected to the noise spectrum of the ligand-receptor binding fluctuations. We analytically derive the bit error probability~(BEP) of the FDD technique, and demonstrate its effectiveness in decoding transmitted concentration signals under stochastic molecular interference, in comparison to a widely-used time-domain detection~(TDD) technique. The proposed FDD method can be applied to any biosensor-based MC-Rxs, which employ receptor molecules as the channel-Rx interface. 
\end{abstract}

\begin{IEEEkeywords}
Molecular communications, receiver, frequency-domain detection, biosensor, ligand-receptor interactions 
\end{IEEEkeywords}

\section{Introduction}
Using molecules to encode and transfer information, i.e., Molecular Communications (MC), is nature's way of connecting \emph{bio things}, such as natural cells, with each other. Engineering this unconventional communication paradigm to extend our connectivity to synthetic \emph{bio-nano things}, such as nanobiosensors, artificial cells, is the vision that gave rise to the \emph{Internet of Bio-Nano Things (IoBNT)}, a novel networking framework promising for unprecedented healthcare and environmental applications of bionanotechnology \cite{akan2016fundamentals, akyildiz2020panacea}. 

Being fundamentally different from the conventional electromagnetic communication techniques, MC requires novel transceiver architectures along with new modulation, coding, and detection techniques that can cope with the highly time-varying, nonlinear, and complex channel characteristics in biochemical environments~\cite{kuscu2019transmitter}. 
The design of MC receivers~(MC-Rxs) and detection techniques has unquestionably attracted the most attention in the literature. However, due to the simplicity it provides in modeling, many of the previous studies considered passive Rx architectures, that are physically unlinked from the MC channel, and thus, of little practical relevance \cite{kuscu2019transmitter}. An emerging trend in MC is to model and design more practical MC-Rxs that employ ligand receptors on their surface as selective biorecognition units, resembling the sensing and communication interface of natural cells. One such design, which was practically implemented in \cite{kuscu2021fabrication}, is based on field-effect transistor biosensors (bioFETs), where the ligand-receptor (LR) interactions are translated into electrical signals via field-effect for the decoding of the transmitted information. 

LR interactions are fundamental to the sensing and communication of natural cells. However, the selectivity of biological receptors against their target ligands is not ideal, and this so-called receptor promiscuity results in cross-talk of other types of molecules co-existing in the biochemical environment \cite{mora2015physical}. This cross-talk is often dealt with by natural cells through intracellular chemical reaction networks and multi-state receptor mechanisms, such as kinetic proofreading~\cite{kuscu2019channel}. The same molecular interference problem also applies to abiotic MC-Rxs that employ ligand receptors, and thus, should be addressed in developing reliable detection techniques \cite{kuscu2022detection}. 

Our previous studies on biosynthetic MC-Rxs have addressed the molecular interference problem by developing detection techniques based on sampling the bound time intervals of individual receptors to discriminate between interferer and information molecules \cite{kuscu2022detection, kuscu2019channel}. However, this approach is not plausible for biosensor-based MC-Rxs, which have no access to time-trajectory of individual receptor states. On the other hand, decoding information from the time-varying concentration of bound receptors performs poorly due to the indistinguishability of different ligand types in time-domain, especially when the Rx does not have any knowledge of the statistics of the interferer concentration, and when the Rx operates near saturation \cite{kuscu2022detection}. 

In this paper, we develop a frequency-domain detection~(FDD) technique for biosensor-based MC-Rxs based on LR binding interactions, which can distinguish different types of ligands co-existing in the channel and estimate their individual concentrations from the power spectral density (PSD) of the fluctuations in receptor occupancy, i.e., binding noise. 

Stochastic and reversible LR interactions can be modeled as a two-state continuous-time Markov process at equilibrium where the state transition rates are given by the binding and unbinding rates of LR pair \cite{mora2015physical}. Although many different types of ligands can interact with the same type of receptors, these interactions are typically governed by different binding and unbinding rates. This difference in reaction rates is reflected to a difference in characteristic frequency $f_{ch}$ of the interactions, which is the reciprocal of the correlation time $\tau_B$ of the Markov process at equilibrium, and also a function of ligand concentration and LR reaction rates \cite{kuscu2022detection}. The characteristic frequency of the LR pair manifests itself as a cut-off frequency in the Lorentzian-shaped PSD of the binding noise. The proposed FDD method exploits this correlation in the frequency domain to estimate the concentration of information molecules in a Maximum Likelihood (ML) manner, and using the estimated concentration, it optimally decodes the transmitted information. We obtained the bit error probability (BEP) for FDD in closed form and compared it to the error performance of a time-domain detection (TDD) technique, which relies on the number of bound receptors, sampled at a single sampling point. The results of the performance analysis indicate that the proposed FDD method vastly outperforms the TDD method, especially at high interference conditions.

\section{System Model}
\label{sec:system_mod}

\begin{figure}[!t]
    \centering
  \includegraphics[scale = 0.41]{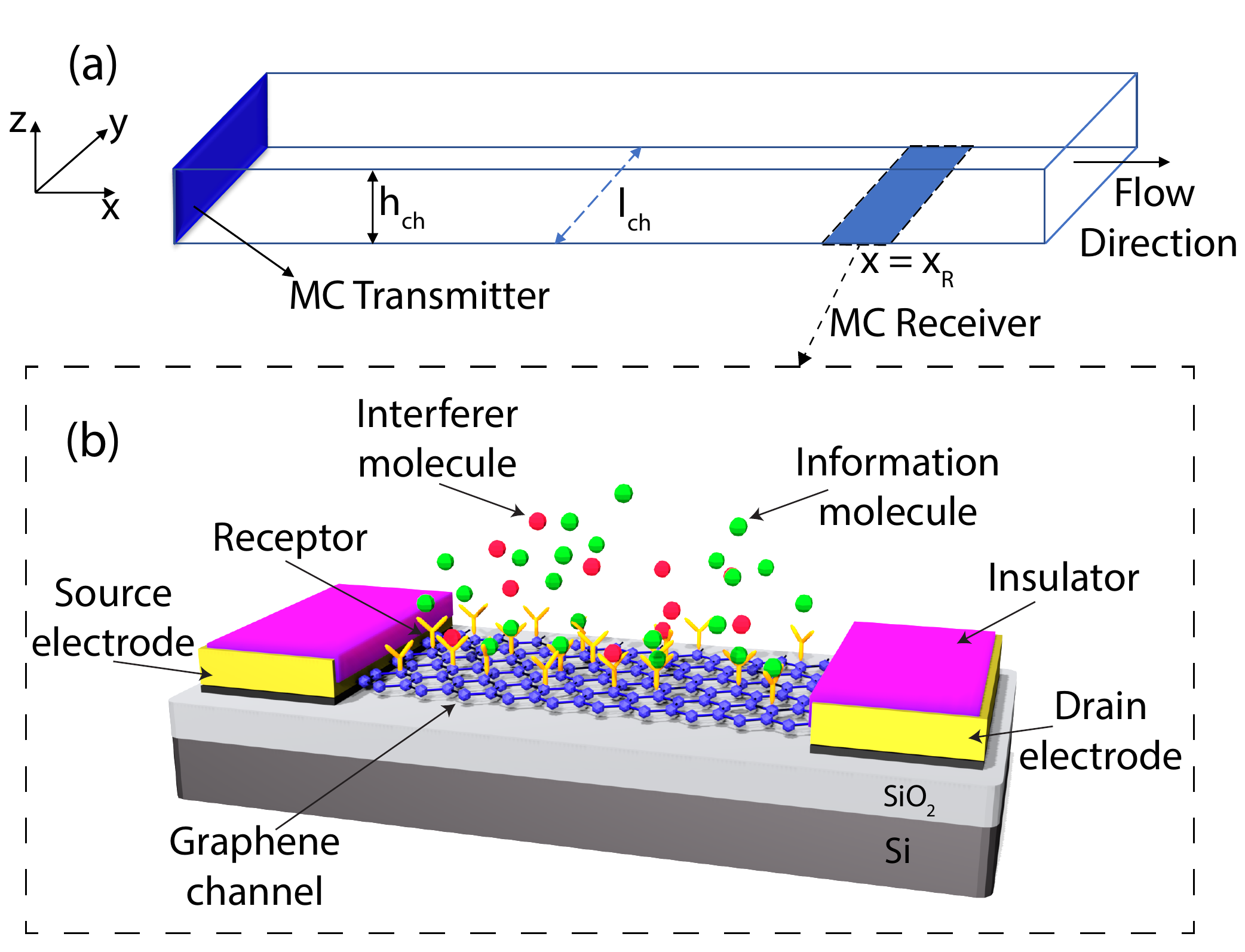}
    \caption{a) 3D view of the microfluidic channel; the locations of Tx and Rx are shown. b) Graphene FET-based MC-Rx exposed to information and interferer molecules.}
    \label{gfet}
\end{figure}

We consider a microfluidic MC system utilizing binary concentration shift keying (CSK) such that the transmitter~(Tx) instantly releases $N_{m|s}$ number of molecules at the beginning of each signaling interval \cite{kuscu2018modeling}. Here $m$ stands for information molecules, and $s\in \{0,1\}$ denotes the transmitted bit. The signaling interval is assumed to be large enough to neglect inter-symbol interference (ISI). The microfluidic channel is abstracted as a 3-dimensional channel with a rectangular cross-section, as shown in Fig. \ref{gfet}(a). Tx is located at the channel inlet, and the molecules are released instantly and uniformly across the cross-section of the channel and propagate through unidirectional fluid flow from Tx to Rx, which is located at the channel bottom. We consider a two-dimensional graphene bioFET-based MC-Rx as illustrated in Fig. \ref{gfet}(b)\cite{kuscu2021fabrication}.  There is a single type of interferer molecules in the channel, which can also bind the receptors on Rx, though with different reaction rates. The concentration of the interferer molecules in the Rx's vicinity, $c_i$, at the sampling time is assumed to follow a log-normal distribution with mean $\mu_{c_i}$ and variance $\sigma^2_{c_i}$. We assume that Rx has the knowledge of the number of information molecules transmitted, $N_{m|s}$, and the binding/unbinding rates of information and interferer molecules. 

The released molecules propagate along the microfluidic channel through convection and diffusion. While convection results in the uniform and unidirectional drift of the transmitted molecules from Tx to Rx, diffusion acts in all directions causing the dispersion of the molecules as they propagate. The dispersion results in a smooth concentration profile which can be approximated by a Gaussian distribution. Assuming that the number of ligands binding the receptors is low enough to neglect the change of concentration in the channel, the propagation can be represented as a one-dimensional convection-diffusion problem with the following solution \cite{kuscu2018modeling}: 
\begin{align}
    c_{m|s}(x,t) = \frac{N_{m|s}}{A_{ch}
\sqrt{4\pi D t}} \exp(-\frac{(x-ut)^2}{4 D t}), 
    \label{c_s}
\end{align}
where $c_{m|s}(x,t)$ is the ligand concentration at position $x$ and time $t$, $A_{ch} = h_{ch} \times l_{ch}$ is the cross-sectional area of the channel with $h_{ch}$ and $l_{ch}$ being the channel height and width, respectively, $u$ is  fluid flow velocity in the x-axis, and $D$ is the effective diffusion coefficient. For channels with rectangular cross-section, $D$ can be expressed as follows \cite{bicen2013system}:
\begin{equation}
    D= \Bigg(1+\frac{8.5 u^2 h_{ch}^2 l_{ch}^2}{210 D_{0}^2(h_{ch}^2+2.4h_{ch}l_{ch}+l_{ch}^2)}\Bigg)D_0,
\end{equation}
where $D_0$ is the diffusion coefficient of the ligand.

The peak of ligand concentration profile given by \eqref{c_s} 
reaches the Rx's center position, $x_R$, at time $t_D=\frac{x_R}{u}$. As the MC channel characteristic is similar to a low-pass filter due to diffusion, the concentration signal is slowly varying around the Rx position, thus allowing equilibrium conditions for the LR reactions with steady ligand concentration in a short time window around $t_D$ \cite{bicen2013system, kuscu2018modeling}. Rx can sample the receptor states at time $t = t_D$ when the ligand concentration is $c_{m|s}(x_R,t_D)= \frac{N_{m|s}}{A_{ch}\sqrt{4\pi Dt_D}}$ \cite{kuscu2016modeling}. Therefore, the number of bound receptors, $N_{b|s}$, follows Binomial distribution with mean $\mu_{N_{b|s}}= p_{b|s} N_r$ and variance $\sigma_{N_{b|s}}^2 = p_{b|s} (1-p_{b|s}) N_r$ \cite{kuscu2016modeling}, where 
$N_r$ is the number of independent surface receptors. Bound state probability of a single receptor, $p_{b|s}$, in the presence of two different types of ligands, i.e., information and interferer molecules,  is given as \cite{kuscu2019channel}
\begin{align}
   p_{b|s}= \frac{c_{m|s}/K_{D_m} + c_{i}/K_{D_i}}{1 + c_{m|s}/K_{D_m} + c_{i}/K_{D_i}}, 
\end{align}  
where $K_{D_m} = k^{-}_m/k^{+}_m$ and $K_{D_i} = k^{-}_i/k^{+}_i$ is the dissociation constant of information and interferer molecules, respectively. The binding of charged ligands to the receptors creates an effective charge reflected on the graphene channel as expressed by 
$Q_{Gr|s} = N_{b|s} q_{eff} N_{e^{-}}$, where 
$N_{e^{-}}$ is number of free electrons per ligand molecules. $q_{eff}$ is the effective charge of a single electron of a bound ligand in the presence of ionic screening, i.e., Debye screening: $q_{eff}= q \times \exp(-\frac{r}{\lambda_D})$, where $q$ is the elementary charge, $r$ is the length of a surface receptor, and $\lambda_D$ is Debye length whose relation is given by $\lambda_D = \sqrt{(\epsilon \kappa_B T)/(2 N_A q^2 c_{ion})}$, where $\epsilon$ is the permittivity of the medium, $\kappa_B$ is the Boltzmann's constant and $N_A$ is the Avogadro's constant \cite{kuscu2016modeling}. Then, the mean surface potential due to bound molecules can be written as $\Psi_{Gr|s}= \frac{Q_{Gr|s}}{C_{G}}$, where $\squeezespaces{0.2}C_{G}= \left(\frac{1}{C_{Gr}}+\frac{1}{C_Q}\right)^{-1}$ is the total gate capacitance of the bioFET. $C_{Gr}$ is the electrical double layer capacitance between graphene and electrolyte 
channel, $C_{Gr}= A_{Gr}\epsilon/\lambda_D$, with $A_{Gr}$ being the area of graphene surface exposed to the electrolyte, and $C_Q$ is quantum capacitance, $C_Q = c_q \times A_{Gr}$, where $c_q$ is the quantum capacitance of graphene per unit area \cite{kuscu2021fabrication}. The deviation in the  output current due to bound molecules at equilibrium is
\begin{equation}
    \Delta I_{b|s} = g \times \Psi_{Gr|s}, 
\end{equation}
where $g$ is the bioFET transconductance. For large $N_r$, the number of bound receptors $N_{b|s}$ at the sampling time can be approximated as Gaussian distributed \cite{kuscu2016modeling}, i.e., $N_{b|s} \sim \mathcal{N}(\mu_{N_{b|s}},\sigma^2_{N_{b|s}})$. As the transduction process is linear, the change in the output current due to bound molecules can also be approximated as Gaussian with mean $\mu_{\Delta I_{b|s}} = \zeta\mu_{N_{b|s}}$ and variance $\sigma^2_{\Delta I_{b|s}} = \zeta^2 \sigma^2_{N_{b|s}}$, where $\zeta = \left(\frac{q_{eff}  N_{e^{-}}  g}{C_G}\right)$.

Another type of noise that contributes to the overall output current fluctuations in low-dimensional semiconductor materials is $1/f$ noise, which depends on the gate voltage and is independent of the received signal. We use the commonly utilized charge-noise model describing the behavior of $1/f$ noise in graphene FETs \cite{heller2010charge}:
    $S_{f}(f) = S_{f_{1Hz}}/f^\beta$
where $S_{f_{1Hz}}$ is the noise power  at 1 Hz, and the noise exponent $\beta$ is an empirical parameter $0.8\leq\beta\leq1.2$. As discussed in \cite{kuscu2016modeling}, $1/f$ noise can be approximated as white noise within physically relevant observation windows. Based on this, the variance of $1/f$ noise can be written as
\begin{align}
    \sigma^2_{f}= \int_{0}^{f_L} S_{f}(f_L) \mathrm{d}f + \int_{f_L}^{f_H} S_{f}(f) \mathrm{d}f,
\end{align}
where $f_L$ is the lower frequency of the observation window, below which the noise power is considered constant, and $f_H$ is the upper frequency, beyond which the noise power is assumed to be negligible. Hence, the variance and mean of total output current variance is $
    \sigma^2_{\Delta I_s} = \zeta^2 \sigma^2_{N_{b|s}} + \sigma^2_{f}$ and  $\mu_{\Delta I_s} = \mu_{\Delta I_{b|s}}$.

\section{Time-Domain Detection}
Since Rx has no knowledge of the interferer concentration statistics, it constructs the optimal ML decision threshold for TDD solely based on its knowledge of the received signal statistics corresponding to the transmitted concentration of information molecules  \cite{kuscu2022detection}:
\begin{equation}
\begin{aligned}
&\gamma_{td} = 
\frac{1}{\sigma_{\Delta I_1}^2-\sigma^2_{\Delta I_0}}\bigg(\sigma^2_{\Delta I_1} \mu_{\Delta I_0} - 
\sigma^2_{\Delta I_0} \mu_{\Delta I_1} +  \sigma_{\Delta I_1} \sigma_{\Delta I_0}\\ & \times \sqrt{(\mu_{\Delta I_1} - \mu_{\Delta I_0})^2 + 2(\sigma^2_{\Delta I_1} -\sigma^2_{\Delta I_0}) \ln(\sigma_{\Delta I_1}/\sigma_{\Delta I_0})} \bigg).
\end{aligned}
\label{th_td}
\end{equation}
As Rx does not account for interference statistics in calculating $\gamma_{td}$, it uses the bound state probability corresponding to a single molecule case, namely, $p_{b|s}= \frac{c_{m|s}/K_{D_m} }{1 + c_{m|s}/K_{D_m}}$.

To derive the BEP for TDD, we first obtain the statistics of the receiver output. By applying the law of total expectation, we can express the mean number of bound receptors as follows:
   $\mu_{N_{b|s}} = \int_{0}^{\infty}  N_r p_{b|s}(c_i) f(c_i)  \mathrm{d}c_i$,       
where  $p_{b|s}(c_i) = \frac{c_{m|s}/K_{D_m} + c_i/K_{D_i}}{1 + c_{m|s}/K_{D_m} + c_i/K_{D_i}}$, and $f(\cdot)$ is the probability density function of log-normal distribution. Hence, $\mu_{\Delta I_s} = \zeta \mu_{N_{b|s}}$. Similarly, by applying the law of total variance, we obtain the output current variance as 
\begin{equation}
\begin{aligned}
  &\sigma^2_{\Delta I_s} = \zeta^2 \bigg(\int_{0}^{\infty} \left(1-p_{b|s}(c_i)\right) p_{b|s}(c_i) N_r f(c_i) \mathrm{d}c_i \\
  & + \int_{0}^{\infty} \left(p_{b|s}(c_i) N_r\right)^2 f(c_i) \mathrm{d}c_i \bigg)  - \mu_{\Delta I_s}^2 + \sigma^2_{f}.
  \end{aligned}
\end{equation}
Therefore, given the decision threshold $\gamma_{td}$, BEP for time detection method can be expressed as follows~\cite{kuscu2022detection}:
\begin{equation}
   P^{TDD}_e = \frac{1}{4} \erfc \left(\frac{\gamma_{td} - \mu_{\Delta I_0}}{\sqrt{2 \sigma^2_{\Delta I_0}}} \right)  +
\frac{1}{4}\erfc \left(\frac{\mu_{\Delta I_1} - \gamma_{td}}{\sqrt{2 \sigma^2_{\Delta I_1}}}\right). 
\end{equation}

\section{Frequency-domain Detection}
In this section, we introduce the FDD method utilizing the model and observed PSD  of the overall noise process (binding noise $+$ $1/f$ noise of the graphene bioFET-based MC-Rx) to estimate the received concentration of information molecules $c_m$, which will be used in symbol decision. Here, the observed PSD is the periodogram  of the noise constructed with the time-domain samples. In the sequel, we  describe the model PSD and then introduce the proposed estimation method.  

\subsection{Theoretical Model of Binding Noise PSD}
This section describes the theoretical model of the binding noise PSD for a particular pair of information and interference concentration, namely $\bm{\lambda} = [c_m, c_i]$.
The binding process of receptors can be described by the Langmuir reaction model with three states, i.e., unbound (R), bound with information molecules (RM) and bound with interferer molecules (RI), with state occupation probabilities $p_R, p_{RM}$ and $p_{RI}$, respectively \cite{mele2020general}:
    $ R + M \underset{k_m^+}{\stackrel{k_m^-}{\rightleftharpoons}} RM,
     \text{and }R + I \underset{k_i^+}{\stackrel{k_i^-}{\rightleftharpoons}} RI$.
Hence, the chemical master equations are expressed as follows: 
\begin{equation}
\begin{bmatrix}
\dfrac{\mathrm{d}p_{RM}}{\mathrm{d}t}\\
\dfrac{\mathrm{d}p_{RI}}{\mathrm{d}t}\\
\dfrac{\mathrm{d}p_{R}}{\mathrm{d}t}\\
\end{bmatrix} = 
  \begin{bmatrix}
-k_m^- & 0 & k_m^+ c_{m} \\
0 & -k_i^- & k_i^+ c_{i} \\
k_m^- & k_i^- & -k_m^+ c_{m} -  k_i^+ c_{i}
\end{bmatrix}  
\begin{bmatrix}
p_{RM}\\
p_{RI}\\
p_{R}\\
\end{bmatrix} 
\label{eqn3}
\end{equation} 
The matrix containing reaction rates and the concentrations in \eqref{eqn3}, has rank 2 since one state probability can be written in terms of the other two state occupation probabilities as 
  $p_{R} + p_{RM} + p_{RI} = 1. $
Therefore, by setting the left-hand side in \eqref{eqn3} to zero 
the equilibrium probabilities can be obtained as
\begin{equation}
   p_{RM}^0 = 
 \frac{c_{m}/K_{D_m}}{1 + \frac{c_{m}}{K_{D_m}} + \frac{c_i}{K_{D_i}}},
    p_{RI}^0 = 
 \frac{c_i/K_{D_i}}{1 + \frac{c_{m}}{K_{D_m}} + \frac{c_i}{K_{D_i}}}
\end{equation}
and $p_{R}^0 = 1-(p_{RM}^0 + p_{RI}^0).$ In the equilibrium conditions, the state occupation probabilities can be expressed in terms of the equilibrium state probability and the fluctuations around this probability \cite{mele2020general, mucksch2018quantifying} as
\begin{equation}
     p_{j}(t) = p_{j}^{0} + \Delta p_{j}(t), \quad j \in \{RM, RI, R\}.
     \label{eqn_fluc}
\end{equation}
Putting \eqref{eqn_fluc} into \eqref{eqn3} and using Taylor's expansion, the state fluctuations can be expressed as follows  \cite{mele2020general}: 
\begin{equation}
    \frac{\mathrm{d}\Delta \vect{p'}(t)}{\mathrm{d}t} = \matr{\Omega}  \Delta \vect{p'}(t).
    \label{eqnstates}
\end{equation}
In \eqref{eqnstates}, $\Delta \vect{p'}(t) = [\Delta p_{RM}(t); \Delta p_{RI}(t)]$ is the reduced form of the vector containing the state occupation probabilities,
where $\matr{\Omega}$ is 
\begin{equation}
\matr{\Omega} = 
    \begin{bmatrix}
    -k_m^+ c_{{m}}-k_m^- & -k_m^+ \\
    -k_i^+ c_{i} & -k_i^+ c_{i} - k_i^-
    \end{bmatrix}.
\end{equation}
The deviation in the output current of the MC-Rx due to stochastic binding reactions, i.e., $\Delta I_b(t)$, is then obtained as
\begin{equation}
    \Delta I_b(t) =  \frac{q_{eff}~g}{C_G} ~\vect{z}^T \matr{R} \Delta \vect{p'}(t)
    \label{eqncurrent}
\end{equation}
where $\vect{z} = [N_{e^{-}}; N_{e^{-}}; 0]$ is the vector containing the number of elementary charges corresponding to each state and $\matr{R}$ is the transformation matrix such that $\Delta \vect{p}(t) = \matr{R} \Delta \vect{p'}(t).$ As $\Delta I_b(t)$ is a stationary process, the theoretical PSD of the binding noise fluctuations can be found by setting $t= 0$ as follows~\cite{mele2020general}:
\begin{align}\label{mod_PSD}
      &S_b(f) = 2~\mathcal{F} \{\EX [\Delta I_b(t) \Delta I_b(t + \tau)]\} \\ \nonumber
     &= 2~\mathcal{F} \{\EX [\Delta I_b(0) \Delta I_b(\tau)]\} \\ \nonumber
      &=4 N_r \left(\frac{q_{eff} g}{C_G}\right)^2 \vect{z}^\intercal \matr{R}
    \Gamma \left(\Re \{(j 2\pi f \matr{I}_{2\times 2} - \matr{\Omega})^{-1}\}\right)^\intercal \matr{R}^\intercal \vect{z} 
\end{align}
where $\mathcal{F}\{\cdot\}$ stands for Fourier transform, $\matr{I}_{2\times 2}$ is the identity matrix and $\Gamma$ is the matrix containing the expected state probabilities, which is given as follows \cite{mele2020general}: 
\begin{equation}
\Gamma = 
    \begin{bmatrix}
     p_{RM}^0 \left(1 - p_{RM}^0\right) & -p_{RM}^0 p_{RI}^0 \\
    -p_{RM}^0 p_{RI}^0 & p_{RI}^0 \left(1 - p_{RI}^0\right)
    \end{bmatrix}.
\end{equation}
Therefore, the theoretical PSD of the total current noise corresponding to a particular ($c_m, c_i$) pair can be written as 
\begin{equation}
    S(f) = S_b(f) + S_{f}(f).
    \label{spec}
\end{equation}

\subsection{Maximum Likelihood Estimation of PSD Parameters}
\label{MLE}
In the following part, we describe the parameter value extraction, namely the estimation of information and interfering molecule concentrations, $\bm{\lambda} = [c_m, c_i]$, from the noise PSD. The detector uses the estimated information molecule concentration $\hat{c}_m$ for symbol decision, as will be explained in the following section, Sec.~\ref{sec:detection}. Our analysis is based on the following assumptions:
\begin{itemize}
   \item The total noise process, namely the binding fluctuations combined with $1/f$ noise, is stationary, zero-mean with a single-sided spectrum.
   \item Rx is given the model PSD function expressed by \eqref{spec}, and the binding/unbinding rates of information 
and interferer molecules. Rx also has the knowledge of the number of information molecules 
transmitted for bits $s=0$ and $s=1$ as mentioned in Sec. \ref{sec:system_mod}. Therefore, Rx will estimate the steady 
information and interferer concentrations by taking time samples from the output current $\Delta I_b$ in a sampling window, where we consider a single realization of the interferer concentration $c_i$ following log-normal distribution as mentioned in Sec. \ref{sec:system_mod}. The DC component of $\Delta I_b$ is discarded to isolate the noise.
\item 
The information and interferer concentrations are considered constant in the sampling window based on the equilibrium assumption discussed in Sec. \ref{sec:system_mod} \cite{kuscu2016modeling}. 
\item 
 The observed PSD of time domain samples and the parametric model of the PSD expressed by \eqref{spec} will be used in the ML estimation of $\bm{\lambda} = [c_{m}, c_i]$.
 It is assumed that the observed PSD is calculated with the periodogram method. 
\end{itemize}
For each transmitted symbol, we have $N$ number of noise samples $\bm{x} = (x_1, x_2, ..., x_N)$ taken with the sampling period of $\Delta t$. Hence, the total duration of sampling per symbol, namely the length of the sampling window, is $T_d = N \Delta t.$ Periodogram for the sampled signal can be computed from the Discrete Fourier transform~(DFT) of the samples $\bm{x}$.

With even $N$, the periodogram values are then expressed as follows:
    $Y_k = \frac{2\Delta t}{N}|X_k|^2$ where $k = 1,...,N/2-1,$ and $|X_k|$ DFT components of $\bm{x}.$

For a stochastic time series of length $N$, the random variable $W_k = 2\frac{Y_k}{S(f_k)}$ follows chi-squared distribution $\chi^2$ \cite{vaughan2010bayesian}, 
where $S(f_k)$ given by Eq.~\eqref{spec} is the true PSD at frequency $f_k$  and 
$f_k = \frac{k}{ N \Delta t}$ and $k = 1,...,N/2-1.$
The $\chi^2$ distribution with two degrees of freedom is in fact the exponential distribution \cite{barret2012maximum}. Therefore, the periodogram values are exponentially distributed about the true PSD with the following probability given the model PSD value at a given frequency:  
\begin{equation}
	p(Y_k|S(f_k)) = \frac{1}{S(f_k)} \mathrm{e}^{-\frac{Y_k}{S(f_k)}},
	\label{eqn_period}
\end{equation}
following that $S(f_k)$ is also expectation value at $f_k$
\cite{barret2012maximum}. Based on \eqref{eqn_period}, 
the likelihood of observing a pair of particular information and interferer concentrations, $\bm{\lambda} = [c_{m}, c_i]$, is
\begin{equation}
\begin{aligned}
\mathcal{L}(\bm{\lambda}) 
	= 
	\prod_{k = 1}^{N/2-1}p(Y_k|S(f_k,\bm{\lambda})) 
	&= \prod_{k = 1}^{N/2-1} \frac{1}{S(f_k,\bm{\lambda})} \mathrm{e}^{-\frac{Y_k}{S(f_k,\bm{\lambda})}},
\end{aligned}
\end{equation}
where $\bm{\lambda} = [c_m,c_{i}]$ is the parameters to be estimated. Here, we use Whittle likelihood, which can be a good approximation to the exact likelihood asymptotically, and also provide computational efficiency, i.e., $O(n\log n)$ compared to $O(n^2)$ for exact likelihood \cite{barret2012maximum, sykulski2019debiased}. Accordingly, the quasi-log likelihood can be written as follows:
\begin{equation}
 \ln\mathcal{L}(\bm{\lambda}) = -\sum_{k = 1}^{N/2-1}\bigg(\frac{Y_k}{S(f_k,\bm{\lambda})} + \ln S(f_k,\bm{\lambda})\bigg).
 \label{eqn7}
\end{equation}
ML estimator extracts the value of $\bm{\lambda}$, i.e., $\hat{\bm{\lambda}}$, that maximizes \eqref{eqn7}. Maximizing $\ln\mathcal{L}(\bm{\lambda})$ is equivalent to minimizing $l = -\ln\mathcal{L}(\bm{\lambda})$ \cite{anderson1990modeling}, such that
\begin{equation}
	\hat{\bm{\lambda}} = \arg\min_{\bm{\lambda}}\{l\}.
\label{eqn8}
\end{equation}
Eq.~\eqref{eqn8} can be solved using numerical methods such as Newton-Ralphson method attaining the ML within few iterations \cite{pfefferle2020whittle}.

\subsection{Symbol Detection}
\label{sec:detection}

\begin{algorithm}[!t]
\caption{Algorithm for the FDD}\label{alg:cap}
\begin{algorithmic}[1]
\State Run Newton method with initial guess $\bm{\lambda}^0 = [c_m^0, c_i^0]$ to find optimal $\bm{\lambda}^{*} = [c_m^{*}, c_i^{*}]$ satisfying \eqref{eqn8}.
\State $\hat{c}_m \gets c_m^{*}$
\State Find the decision threshold $\gamma_{fd}$.
\State Run threshold operation:
     \If {$\hat{c}_m > \gamma_{fd}$} $\text{Estimated bit } \hat{s} \gets 1$
    \Else {} $\text{Estimated bit } \hat{s} \gets 0$
    \EndIf 
\end{algorithmic}
\end{algorithm}

The ML estimator described in Sec.~\ref{MLE} is asymptotically unbiased such that $\hat{\bm{\lambda}}$ tends to have multi-normal distribution \cite{toutain1994maximum} with $\EX[\hat{\bm{\lambda}}] = \bm{\lambda}$, 
and the respective variance of the estimated parameters, which is the diagonal elements of inverse Fisher information matrix~(FIM) $\matr{F}(\bm{\lambda})$, 
\begin{equation}
    \sigma^2_{\hat{\lambda_i}} =  (\matr{F}(\bm{\lambda}))^{-1}_{(ii)}, \quad \matr{F}(\bm{\lambda})_{(ij)} = \EX\left(\frac{\partial^2 l}{\partial \lambda_i \partial \lambda_j}\right).
    \label{IFI}
\end{equation}
where the expectation is taken with respect to the probability distribution of the observed spectrum $p(Y_1,Y_2,...,Y_N)$. 
Putting $\squeezespaces{0.5}l=-\ln\mathcal{L}(\bm{\lambda})$ into \eqref{IFI}, the FIM can be expanded as:
\begin{equation}
\begin{aligned}
    &\matr{F}_{(ij)} = \\ &\EX \left( \sum_{k = 1}^{N/2-1} \frac{S(f_k)-Y_k}{S(f_k)^2} \frac{\partial^2 S}{\partial\lambda_i \partial \lambda_j} + \frac{2 Y_k-S(f_k)}{S(f_k)^3} \frac{\partial S}{\partial\lambda_i} \frac{\partial S}{\partial\lambda_j}\right).
\end{aligned}
    \label{FI}
\end{equation}
Considering that $S(f)$ is a slowly varying function, there is no need to calculate individual periodogram values in \eqref{FI}. Because periodogram values can be smoothed by summing over frequency such that $\sum_{n=1}^{N/2-1} Y_k \phi(f_k) \simeq \sum_{n=1}^{N/2-1} S(f_k) \phi(f_k),$ for any smooth function $\phi(f_k)$ ~\cite{levin1965power,toutain1994maximum}. Based on this, Eq.~\eqref{FI} can be simplified as~\cite{toutain1994maximum}
\begin{equation}
    \matr{F}_{(ij)} \simeq \sum_{k=1}^{N/2-1} \frac{1}{S(f_k)^2} \frac{\partial S}{\partial \lambda_i}\frac{\partial S}{\partial \lambda_j},
    \label{fisher}
\end{equation}
where the derivatives are taken at the true value of the parameters. This is a good approximation for the large number of samples such that 
periodogram values can be approximated as Gaussian by the central limit theorem \cite{libbrecht1992ultimate}.
Rx decides the transmitted bit by applying the ML decision rule on the estimated information molecule concentration $\hat{c}_m$, as described by the pseudo-algorithm for FDD in Algorithm~\ref{alg:cap}. The ML decision threshold for FDD is 
\begin{equation}
\begin{aligned}
&\gamma_{fd} = 
\frac{1}{ \sigma^2_{\hat{c}_{m|1}} -\sigma^2_{\hat{c}_{m|0}}} 
\bigg(\sigma^2_{\hat{c}_{m|1}} c_{m|0} - 
\sigma^2_{\hat{c}_{m|0}} c_{m|1} +  \sigma_{\hat{c}_{m|1}} \sigma_{\hat{c}_{m|0}}\\ & \times \sqrt{(c_{m|1} - c_{m|0})^2 + 2(\sigma^2_{\hat{c}_{m|1}} -\sigma^2_{\hat{c}_{m|0}}) \ln(\sigma_{\hat{c}_{m|1}}/\sigma_{\hat{c}_{m|0}})} \bigg),
\end{aligned}
\label{FD_threshold}
\end{equation}
where $\sigma^2_{\hat{c}_{m|s}}$ is the variance and $c_{m|s}$ is the expected value of estimated information molecule when the transmitted bit is $s \in \{0,1\}$. 
Since Rx does not know the true value of the interfering molecule concentration, it computes the decision threshold as if there was no interference. 
Therefore, the following model PSD is used while computing the threshold $\gamma_{fd}$: 
\begin{equation}
    S(f,c_m) = 4 N_r \zeta^2 \frac{1}{2\pi f + 1/\tau_{m}} p_{b} (1- p_{b})  +  S_{f}(f),
    \label{no_interference_model}
\end{equation}
where $\tau_{m} = 1/(c_{m} k^{+}_{m} + k^{-}_{m})$ and $p_{b} = \displaystyle\frac{c_{m}}{K_{D_m} + c_{m}}.$
Hence, using \eqref{no_interference_model}, and \eqref{fisher} for $\bm{\lambda} = [c_m]$, the variance of the estimated information molecule concentration corresponding to the transmitted bit $s \in \{0,1\}$ can be written as $\sigma^2_{\hat{c}_{m|s}} = 1/\sum_{k=1}^{N/2-1} \frac{1}{S(f_k)^2} (\frac{\partial S}{\partial c_m})^2\Bigr|_{\substack{c_m = c_{m|s}}}.$ Note that here the expression for  $\sigma^2_{\hat{c}_{m|s}}$ does not give the actual asymptotic variances since Rx estimates the value of $c_m$ based on the model PSD described by \eqref{spec}. 
\subsection{Asymptotic Bit Error Probability}
To calculate BEP for FDD, we need the actual values of the variance of estimated information molecule concentration corresponding to $\squeezespaces{0.5}s=0\text{ and }s = 1$, i.e., $\sigma^2_{\hat{c}_{m|s}}$. Using the model PSD $S(f,(c_m,c_i))$ given by \eqref{spec}, and \eqref{fisher} with $\bm{\lambda} = [c_m,c_i]$, 
the variance can be expressed as $\squeezespaces{0.2}\sigma^2_{\hat{c}_{m|s}}=(\matr{F}_s(\bm{\lambda}))_{(11)}^{-1}$, where $\matr{F}_s$'s elements are: 
\begin{align}
 \matr{F}_{{s}_{(11)}} &=  \sum_{k=1}^{N/2-1} \frac{1}{S(f_k)^2} \left(\frac{\partial S}{\partial c_m}\right)^2 \Bigr|_{\substack{c_m = c_{m|s} \\c_i = \mu_{c_i}}},\\
\matr{F}_{s_{(22)}} &=  \sum_{k=1}^{N/2-1} \frac{1}{S(f_k)^2} \left(\frac{\partial S}{\partial c_i}\right)^2 \Bigr|_{\substack{c_m = c_{m|s} \\c_i = \mu_{c_i}}},\\
\matr{F}_{s_{(12),(21)}} &=   \sum_{k=1}^{N/2-1} \frac{1}{S(f_k)^2} \left(\frac{\partial S}{\partial c_m}\right) \left(\frac{\partial S}{\partial c_i}\right) \Bigr|_{\substack{c_m = c_{m|s} \\c_i = \mu_{c_i}}}.
\end{align}   
As a result, BEP for FDD can be written as 
\begin{equation}
P^{FDD}_e = \frac{1}{4} \erfc \left(\frac{\gamma_{fd} - c_{m|0}}{\sqrt{2 \sigma^2_{\hat{c}_{m|0}}}} \right) +
\frac{1}{4}\erfc \left(\frac{c_{m|1} - \gamma_{fd}}{\sqrt{2 \sigma^2_{\hat{c}_{m|1}}}}\right).
\label{eq:bep_fd}
\end{equation}
Here, it should be noted that \eqref{eq:bep_fd} is an asymptotic expression based on the Gaussian distribution assumption in Sec.~\ref{sec:detection}.

\section{Performance Evaluation}
\label{sec:eval}
In this section, we analyze the performance of FDD and TDD in terms of BEP. The default values of the system parameters are given in Table~\ref{tab:parameters}, with the reaction rates adopted from \cite{kuscu2022detection}. In the rest of the paper, the saturation and the non-saturation corresponds to 
the Rx's receptors being saturated due to high ligand concentrations and far from the saturation, respectively.
To simulate the saturation, the number of transmitted information molecules is taken as $N_{m|s\in\{0,1\}} = [2,5]\times 10^4$. Otherwise, default values are used. 

\begin{table}[!b]\scriptsize		
\caption{Default Values of System Parameters}
\label{tab:parameters}
\centering
\begin{tabular}{m{4.9cm}|m{3cm}}
\hline 
 Temperature~($T$) & $300$K \\ \hline
 Microfluidic channel height ($h_{ch}$), width ($l_{ch}$) & 5 $\mu$m, 10 $\mu$m  \\ \hline
Average flow velocity ($u$) & 10 $\mu$m/s  \\ \hline
 Distance of Rx's center position to Tx ($x_R$) & $1$mm \\ \hline
 Ionic concentration of medium ($c_{ion}$) & 30 mol/m$^3$  \\ \hline
Relative permittivity of medium ($\epsilon/\epsilon_0$) & $80$  \\ \hline
 Intrinsic diffusion coefficient ($D_0$) & $2 \times 10^{-11}$ m$^2$/s  \\ \hline
Binding rate of information and interferer molecules ($k^{+}_m, k^{+}_i$) & $4 \times 10^{-17}$ m$^3$/s \\ \hline
Unbinding rate of information molecules ($k^{-}_m$) & 2 s$^{-1}$  \\ \hline
Unbinding rate of interferers ($k^{-}_i$) & 8 s$^{-1}$  \\ \hline
Average \# of electrons in a ligand ($N_{e^{-}}$) & 3  \\ \hline
Number of independent receptors ($N_r$) & $120$  \\ \hline
Length of a surface receptor ($r$) & 2 nm  \\ \hline
Transconductance of graphene bioFET ($g$)& $1.9044\times 10^{-4}$ A/V \\ \hline
Width of graphene in transistor ($l_{gr}$) & 10$\mu$m  \\ \hline
Quant. capacitance of graphene per unit area 
($c_q$)& $2\times 10^{-2}$ F m$^{-2}$\\ \hline
\# of transmitted ligands for $s = {0,1}$ ($N_{m|s}$) & $[1,5] \times 10^3$  \\ \hline
\# of noise samples ($N$) &  700 \\ \hline
Sampling period ($\Delta t$) & 0.005 s \\ \hline
Mean interference to information concentration ratio ($\gamma = \mu_{c_i}/c_{m|s=1}$) &  1 \\ \hline
Interference mean/std ratio ($\mu_{c_i}/\sigma_{c_i}$) &  10 \\ \hline
Power of $1/f$ noise at 1 Hz ($S_{f_{1Hz}}$) & $10^{-23}$ A$^2/$Hz \\ \hline
\end{tabular}%
\end{table}

\begin{figure}
    \centering
    \includegraphics[width = 0.4\textwidth]{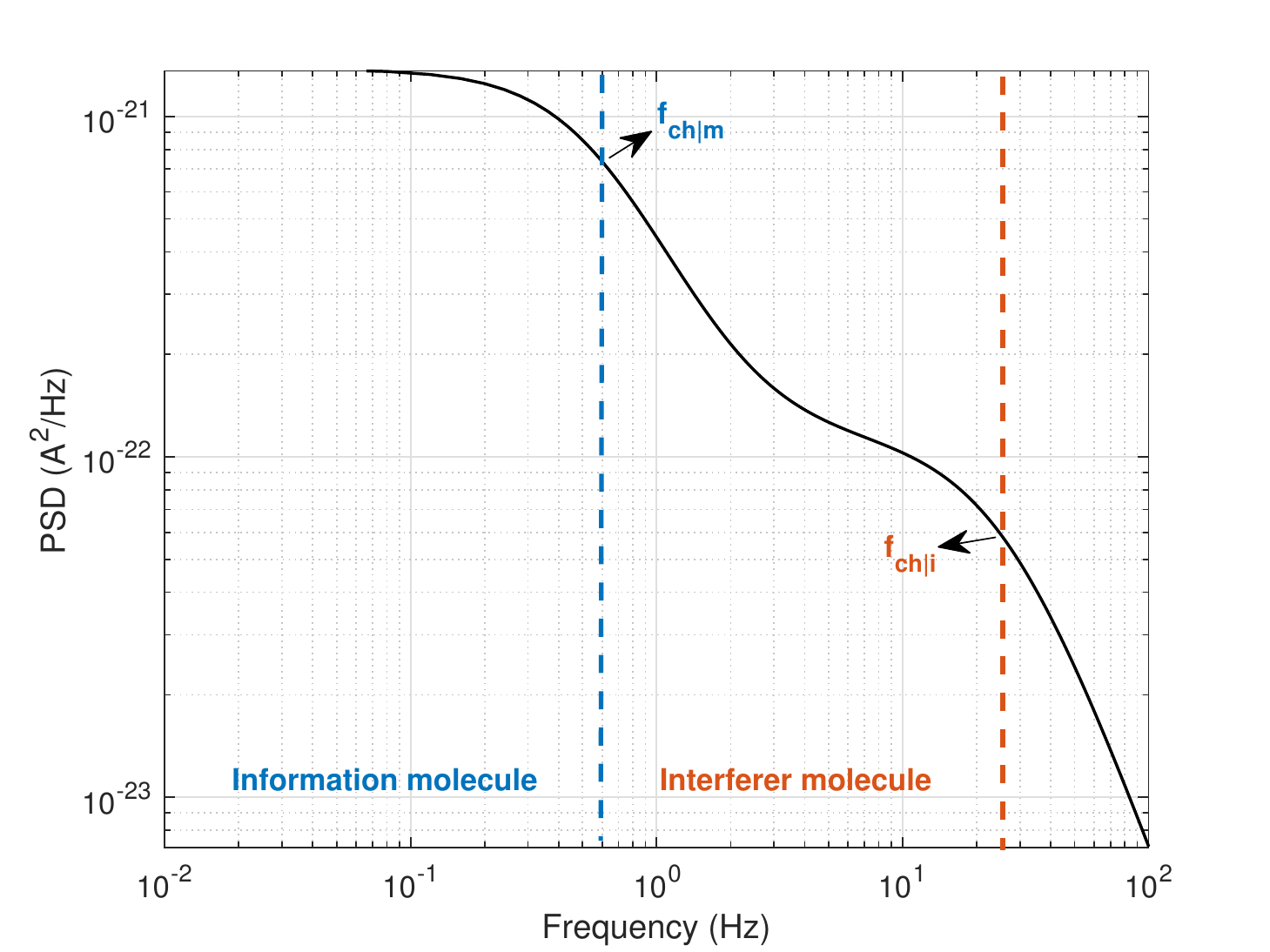}
    \caption{The model PSD with highlighted characteristic frequencies.}
    \label{fig:PSD}
\end{figure}

\begin{figure*}[t]
\centering
\subfloat[]{%
  \includegraphics[width=0.33\textwidth]{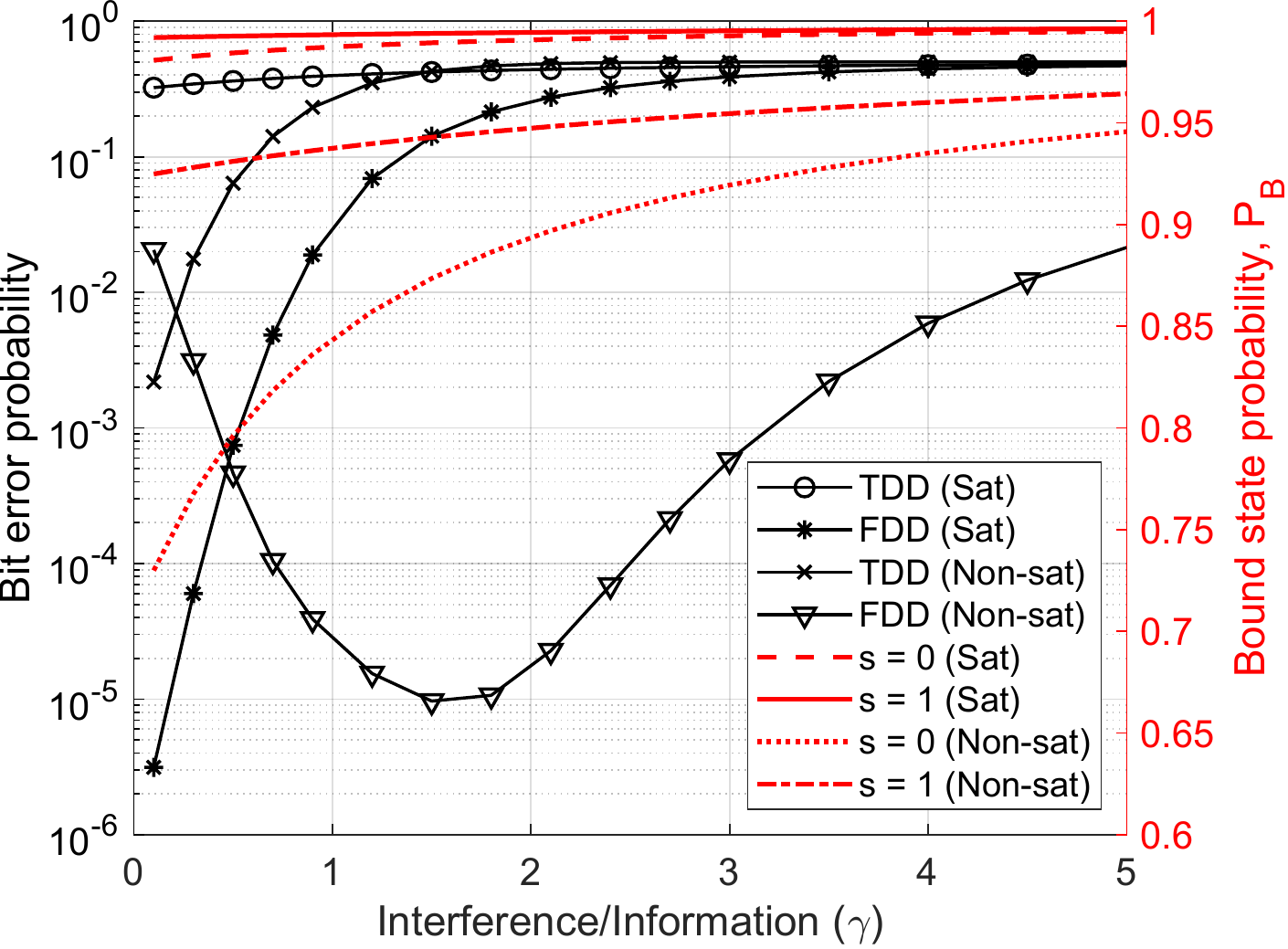}%
  \label{fig:interference}%
}
\subfloat[]{%
  \includegraphics[width=0.33\textwidth]{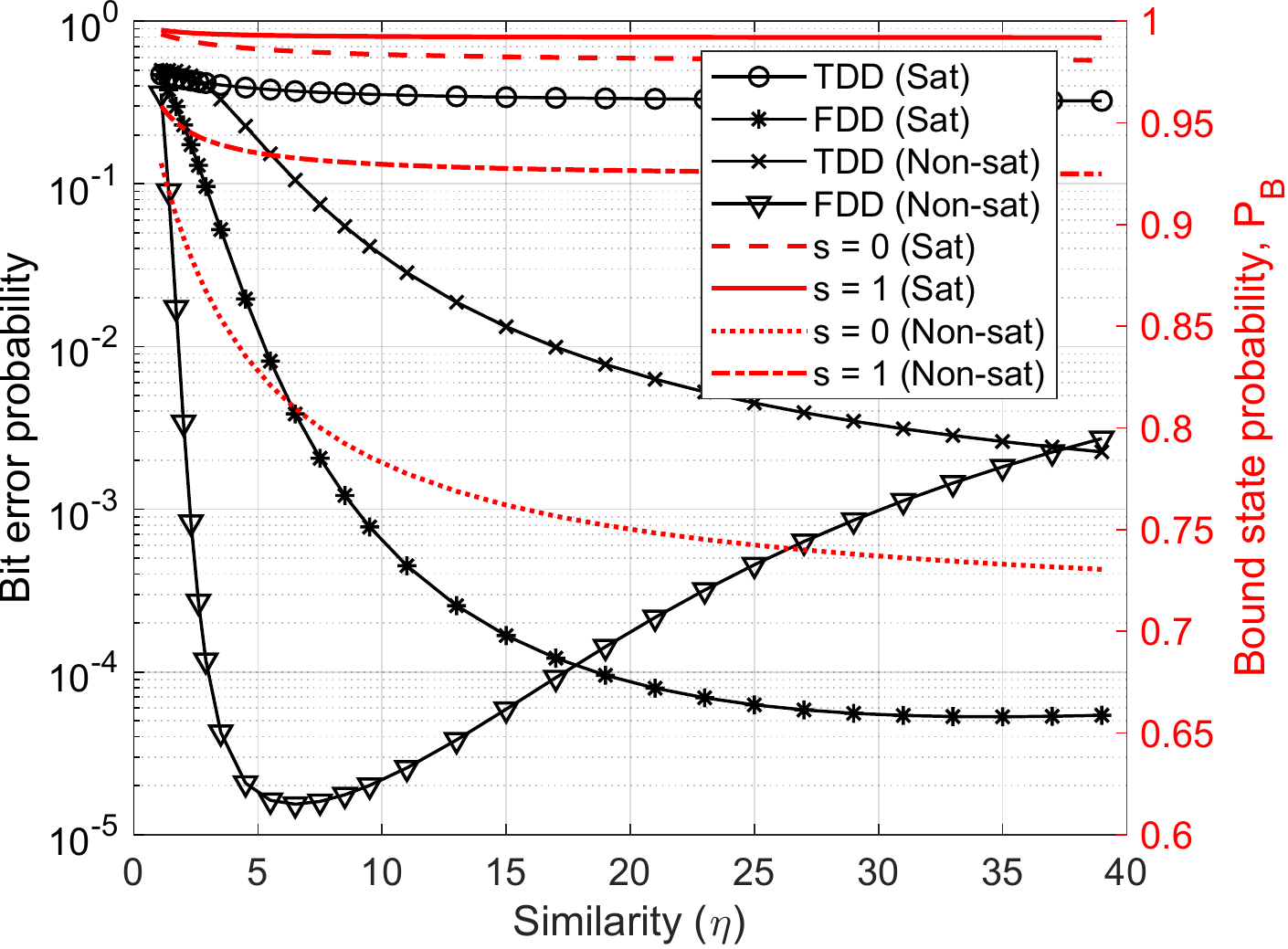}%
  \label{fig:similarity}%
}
\subfloat[]{%
  \includegraphics[width=0.33\textwidth]{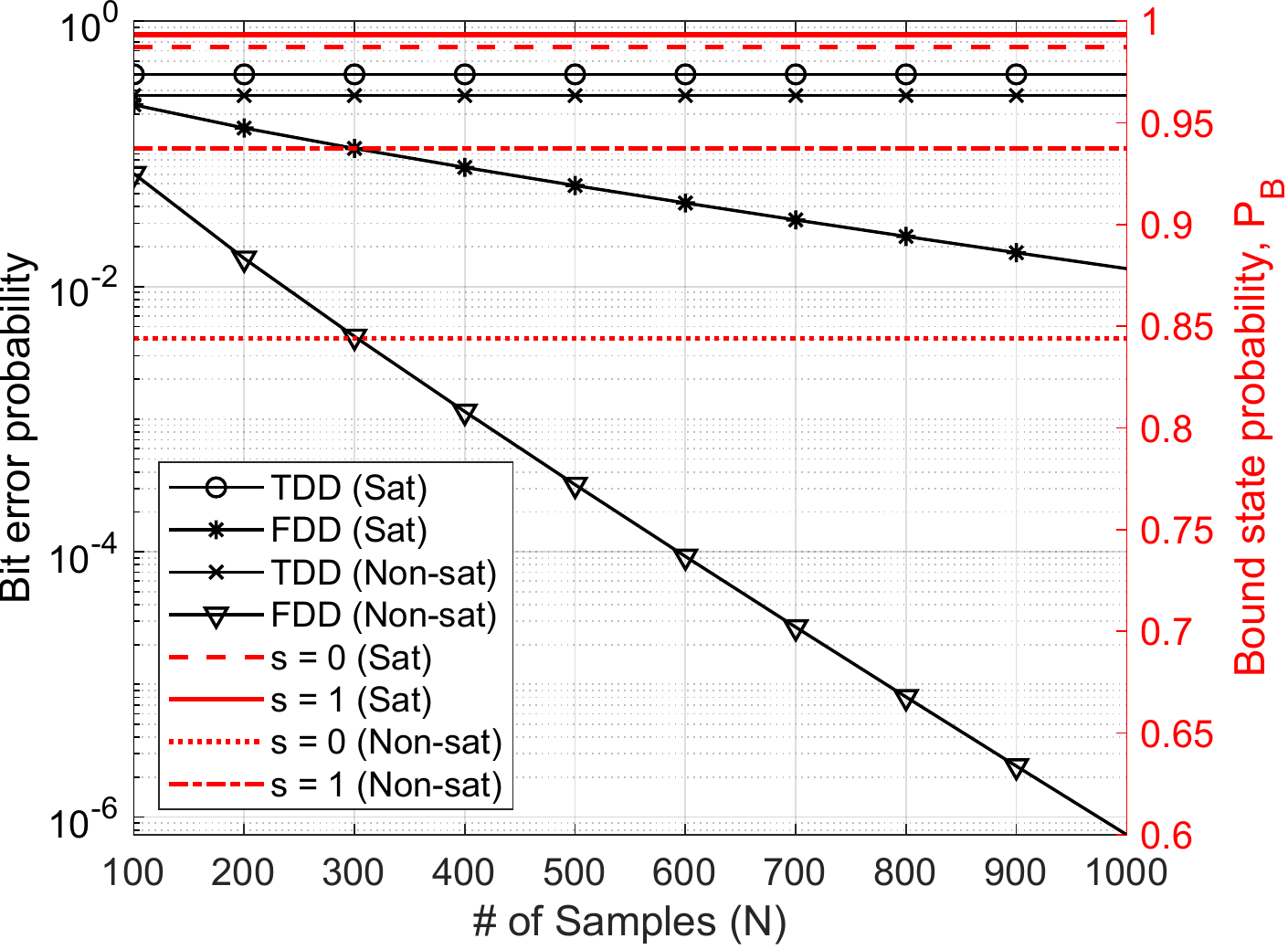}%
  \label{fig:numSamples}%
}
\caption{BEP for varying (a) mean interference concentration level (b) similarity of affinities for information and interferer molecules (c) number of time samples $N$.}
\end{figure*}

We first consider the effect of the mean interference concentration, $\mu_{c_i}$ on the BEP performance of TDD and FDD in saturation and non-saturation conditions. We define a tuning parameter $\gamma$ such that mean interferer concentration is given by $\mu_{c_i} = \gamma~  c_{m|s=1}$. As shown in Fig.~\ref{fig:interference}, FDD outperforms TDD in both scenarios. The performance of TDD degrades dramatically due to Rx saturation with increasing $\mu_{c_i}$. In non-saturation, the performance of FDD improves with increasing $\mu_{c_i}$ up to a certain point, beyond which further increase in $\mu_{c_i}$ degrades the performance of FDD because when Rx is not saturated, the variance of the estimated information molecule concentration $\sigma^2_{\hat{c}_{m|s}}$ is 
minimized at a certain $\mu_{c_i}$ beyond which its value increases with increasing $\mu_{c_i}$. In saturation, however, $\sigma^2_{\hat{c}_{m|s}}$ monotonically increases with $\mu_{c_i}$.  

Next, we consider the effect of similarity parameter,  namely the affinity ratio of information and interferer molecules $\eta = K_{D_i}/K_{D_m}$, on the BEP performance for saturation and non-saturation cases. 
As displayed in Fig.~\ref{fig:similarity}, FDD outperforms TDD in both saturation and non-saturation cases. Regarding the non-saturation case, the performances of both detection methods improve with increasing similarity up to a certain point because the effect of interference on the detection performance weakens; namely, bound state probability for interferer molecules decreases. However, when the similarity is further increased, the performance of FDD degrades. Intuitively, this is because the characteristic frequencies corresponding to bits $s=0$ and $s=1$ \cite{frantlovic2013analysis} 
\begin{equation}
\begin{aligned}
    &\vect{f_{ch|s}} = [f_{ch|m}, f_{ch|i}] \\&= \frac{1}{4\pi}\left[\frac{1}{\tau_{m|s}} + \frac{1}{\tau_{i}} \pm \sqrt{\left(\frac{1}{\tau_{m|s}}-\frac{1}{\tau_{i}}\right)^2 + 4 k_m^{+} c_{m|s} k_i^{+}  c_i }\right],
\end{aligned}
\end{equation}
where $\tau_{m|s} = 1/(c_{m|s} k_m^+ + k_m^-)$
and $\tau_i = 1/(c_i k_i^+ + k_i^-)$, are approaching each other in the spectrum, making it difficult to distinguish the bits. As shown in Fig.~\ref{fig:PSD} for an example scenario, two characteristic frequencies, $f_{ch|m}$ and $f_{ch|i}$, appear in the spectrum for each transmitted bit due to the binding of two types of molecules, namely information and interferer molecules, with the order depending on the concentrations and binding/unbinding rates of the individual molecule types. In the non-saturation case, we do not observe this phenomenon because the characteristic frequencies do not come close to each other to degrade the detection performance with increasing similarity. 
We also consider the effect of the number of time samples, $N$, and the sampling period $\Delta t$ on the BEP performance. As shown in Fig.~\ref{fig:numSamples}, the performance of FDD increases with $N$. This is expected as taking more samples decreases the variance of the estimated information molecule concentration $\sigma^2_{\hat{c}_{m|s}}$, hence, decreases the BEP. For TDD, the performance does not change with $N$ as Rx takes one sample in the sampling window. For varying $\Delta t$, the performance of FDD increases with increasing $\Delta t$ for the non-saturation case. Note that $\Delta t$ should be  shorter than the characteristic time scale of any reactions to be able to capture the fluctuations and to satisfy the sampling at the equilibrium assumption, which is discussed in Sec.~\ref{sec:system_mod}. Therefore, we consider $\Delta t$ values satisfying this condition.

\begin{figure}[t]
\centering

  \includegraphics[width=0.33\textwidth]{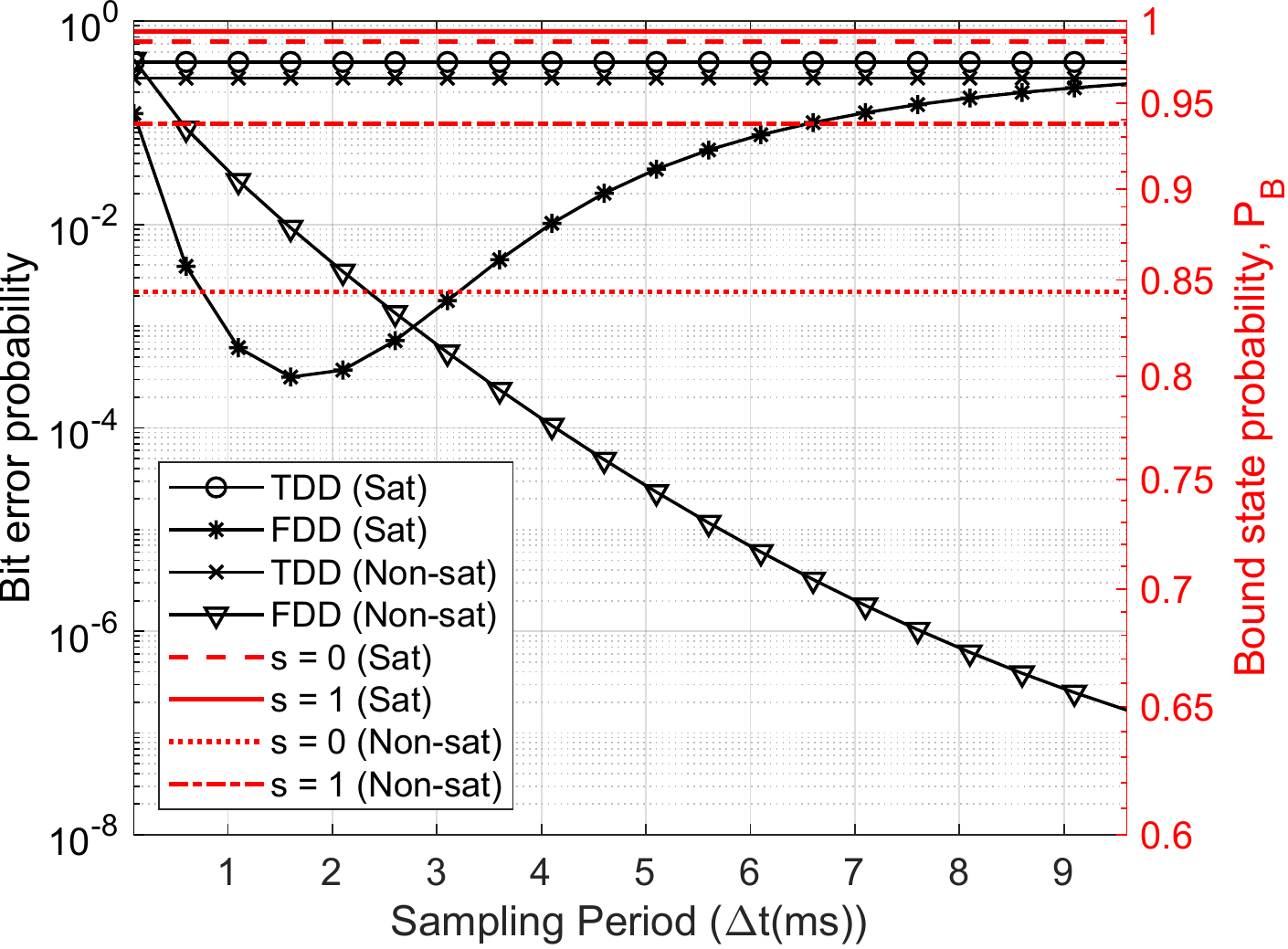}%
  \label{fig:samplingPeriod}%

\caption{BEP for varying 
sampling period.}
\label{fig:4}
\end{figure}

\section{Conclusion}
\label{sec:con}
In this paper, we proposed a FDD method for the FET-based MC-Rx, which utilizes the output noise PSD to extract the transmitted bit. We derived the BEP for the proposed method and a one-shot TDD method considering the existence of a single type of interferer molecules in a microfluidic channel. Our analysis reveals that the proposed detection method significantly outperforms the TDD, primarily when high interference exists in the channel. 
\section*{Acknowledgment}
This work was supported in part by the AXA Research Fund (AXA Chair
for Internet of Everything at Ko\c{c} University), the Horizon 2020 Marie Skłodowska-Curie Individual Fellowship under Grant Agreement 101028935, and by The Scientific and Technological Research Council of Turkey (TUBITAK) under Grant {\#120E301}, and Huawei Graduate Research Scholarship.

\bibliographystyle{IEEEtran}
\bibliography{references}

\vfill

\end{document}